\documentclass[aps,pra,twocolumn,superscriptaddress]{revtex4-1}
\usepackage{color}
\usepackage{dcolumn}
\usepackage{graphicx}

\usepackage{amsmath,amsfonts,amssymb,amsthm}

\usepackage{bbold}
\usepackage{dsfont}

\usepackage{soul}

\usepackage{fancyhdr}
\usepackage{multirow}
\usepackage{bm}
\usepackage{dsfont}
\usepackage{cancel}

\usepackage{xifthen}

\usepackage{float}

\usepackage{graphicx}
\usepackage[usenames,dvipsnames,dvipsnames,svgnames]{xcolor}

\usepackage[colorlinks,bookmarks=false,
linkcolor=blue,
urlcolor=blue,
citecolor=blue
]{hyperref}

\usepackage{xspace}
\newcommand{\ket}[1]{\ensuremath{|#1\rangle}\xspace}
\newcommand{\bra}[1]{\ensuremath{\langle #1|}\xspace}
\def\mean#1{\ensuremath{\langle #1 \rangle}}

\def\dissip#1{\mathcal{D}\big[#1\big]}

\def\commut#1#2{\lbrack #1,#2 \rbrack}

\renewcommand{\H}{\mathcal{H}}

\def\op#1{\hat{#1}}

\renewcommand{\ao}[1][]{%
	\ifthenelse{\equal{#1}{}}{\ensuremath{\op{a}}}{\ensuremath{\op{#1}}}%
}
\newcommand{\co}[1][]{%
	\ifthenelse{\equal{#1}{}}{\ensuremath{\op{a}^{\dagger}}}{\ensuremath{\op{#1}^{\dagger}}}%
}

\usepackage{array,calc}

\begin{document}
\title{A dissipative time crystal in an asymmetric non-linear photonic dimer}

\author{Kilian Seibold}
\affiliation{Institute of Physics, Ecole Polytechnique F\'ed\'erale de Lausanne (EPFL), CH-1015, Lausanne, Switzerland}
\email{kilian.seibold@epfl.ch}
\author{Riccardo Rota}
\affiliation{Institute of Physics, Ecole Polytechnique F\'ed\'erale de Lausanne (EPFL), CH-1015, Lausanne, Switzerland}
\author{Vincenzo Savona}
\affiliation{Institute of Physics, Ecole Polytechnique F\'ed\'erale de Lausanne (EPFL), CH-1015, Lausanne, Switzerland}

\date{\today}

\begin{abstract}
	
	We investigate the behavior of two coupled non-linear photonic cavities, in presence of inhomogeneous coherent driving and local dissipations. By solving numerically the quantum master equation, either by diagonalizing the Liouvillian superoperator or by using the approximated truncated Wigner approach, we extrapolate the properties of the system in a thermodynamic limit of large photon occupation. When the mean field Gross-Pitaevskii equation predicts a unique parametrically unstable steady-state solution, the open quantum many-body system presents highly non-classical properties and its dynamics exhibits the long lived Josephson-like oscillations typical of dissipative time crystals, as indicated by the presence of purely imaginary eigenvalues in the spectrum of the Liouvillian superoperator in the thermodynamic limit. 
		
\end{abstract}

\maketitle

\section{Introduction}
\label{sec:intro}

Open many-body quantum systems \cite{Hartmann16,CiutiRMP,NohAngelakis16} have become a major field of study over the last decade. The open nature is common to a vast class of modern experimental platforms in quantum science and technology, such as photonic systems \cite{Szameit_2010}, ultracold atoms \cite{RevModPhys.80.885, PhysRevA.75.013804, Baumann2010, PhysRevLett.107.140402, Brennecke11763}, optomechanical systems \cite{RevModPhys.86.1391,Pigeau2015,Teufel2011,Kolkowitz1603} or superconducting circuits \cite{Carmichael15, PhysRevX.7.011012, PhysRevX.7.011016}, for which driving and losses are omnipresent. Open quantum systems also display emergent physics, in particular dissipative phase transitions \cite{DallaTorre10, DallaTorre12, Lee13, Sieberer13, Sieberer14, Altman15, Carmichael15, Bartolo16, Mendoza16, CasteelsStorme16, Jin16, Maghrebi16, Marino16, Rota17, Savona17, Casteels17, CasteelsFazio17, FossFeig17, Biondi17, Biella17, Vicentini18, Rota18, Rota19} and topological phases \cite{Lu2014,Roushan2016,St-Jean2017,Umucalilar12,Klembt2018,Dong16}.

Several studies have highlighted the possibility for a continuous-wave driven-dissipative quantum system to reach a non-stationary state in the long time limit in which undamped oscillations arise spontaneously \cite{Iemini18,Wang18,Gong18,Tucker2018,Gambetta19,Tindall19,Lledo19, Buca2019}. This phenomenon has been dubbed as boundary or dissipative time crystal (DTC), in analogy with the time crystals in some Hamiltonian systems \cite{Sacha_2017}. Formally, DTCs are associated with the occurrence of multiple eigenvalues of the Liouvillian with vanishing real and finite imaginary part \cite{Albert14, PhysRevX.6.041031, Baumgartner_2008}. The experimental feasibility of DTC has been confirmed by their observation in phosphorous-doped silicon \cite{OSullivan18}. The research for further platforms showing this phenomenon is very active and important to understand the mechanisms behind the spontaneous breaking of the time-translation symmetry in open quantum many-body systems. 

One of the main difficulties in the realization of DTCs in real system is related to the fragility of this phase to external perturbations which affect the symmetric structure of the model. Indeed, in most of the cases considered so far, the engineering of the DTCs relies on the exploitation of certain symmetries (either manifest \cite{Gong18,Lledo19} or emergent \cite{Gambetta19}) in the Hamiltonian or in the dissipation mechanism, which can be hard to maintain in real driven-dissipative systems out of equilibrium.


In this work, we show that a DTC can arise in a simple system of two coupled photonic cavities, whose equation of motion does not preserve any symmetry but the time-translation invariance. In a broad region of the parameter space, the dynamics of this system presents limit cycles associated to parametric instabilities \cite{Sarchi08}, which can be regarded as the classical limit of a DTC. In this regime the system displays large fluctuations and entanglement, thus departing from its classical analog. As symmetries are not required for the occurrence of a DTC, this system is very robust and may be easily realized for example on a superconducting circuit architecture \cite{Eichler14} or with coupled semiconductor micropillars \cite{Abbarchi2013,Galbiati12,Lagoudakis2010,Rodriguez2016}. This prototypical system is also a minimal model of dissipative Kerr solitons, that are emerging as the most suitable optical system for precision frequency generation and metrology \cite{Kippenbergeaan8083}, and therefore highlights the potential of these devices as sources of strongly nonclassical light. Very recently, the emergence of parametric instabilities in a photonic dimer has been observed in a classical regime of large occupation \cite{zambon2019parametric}.

The paper is organized as follows. In Sec. \ref{sec:theoretical framework} we introduce the quantum model and the theoretical tools used to calculate its properties. In Sec. \ref{sec:results} we discuss the result obtained for both the stationary-state and the dynamics of the system. In Sec. \ref{sec:conclusions}, we draw our conclusions.

\section{Theoretical framework}
\label{sec:theoretical framework}


We consider two coupled Kerr cavities where only one is coherently driven. The system Hamiltonian in a frame rotating with the pump frequency reads (with $\hbar=1$)
\begin{equation}
\begin{split}
\op{\H}=&\sum_{i=1,2} -\Delta \co_i\ao_i + \dfrac{U}{2}\co_i\co_i\ao_i\ao_i\\ 
&- J(\co_1\ao_2+\ao_1\co_2) + F(\co_1+\ao_1)\,,
\end{split}
\label{equa:system Hamiltonian}
\end{equation}
where $\ao_i$ is the bosonic annihilation operator of the $i$-th mode, $\Delta$ is the frequency detuning between the pump and the resonator, $U$ is the on-site interaction strength, $J$ is the hopping coupling and $F$ is the driving amplitude. 
The dissipative dynamics can be described within the Born-Markov approximation, resulting in the following Lindblad quantum master equation \cite{OpenQBreuer,GardinerZollerQNoise} for the density matrix $\op{\rho}$,
\begin{equation}
\frac{d\op{\rho}}{dt}=\mathcal{L}\op{\rho} = -i\commut{\op{\H}}{\op{\rho}}+\sum_{i=1,2}\kappa\dissip{\ao_i}{\op{\rho}}\,.
\label{equa:master equation}
\end{equation}

Here $\mathcal{D}\big[\ao_i\big]\op{\rho} = \ao_i\op{\rho}\ao_i^{\dagger}-1/2(\ao_i^{\dagger}\ao_i\op{\rho}+ \op{\rho}\ao_i^{\dagger}\ao_i)$ is the dissipator in Lindblad form accounting for losses to the environment and $\kappa$ the dissipation rate. $\mathcal{L}$ is the Liouvillian superoperator and its spectrum encodes the full dynamics of the open quantum system. The expectation value of any quantum mechanical observable $\op{o}$ over the state characterized by the density matrix $\op{\rho}$ is computed as $\mean{\op{o}}= \text{Tr}(\op{o}\op{\rho})$. In the long time limit, the system evolves towards a non-equilibrium steady state $\op{\rho}_{ss}$ satisfying the condition $d\op{\rho}_{ss}/dt=0$. We determine the steady-state density matrix by numerically solving the linear system $\mathcal{L}\op{\rho}_{ss}=0$, and imposing the condition $\text{Tr}(\op{\rho}_{ss})=1$. The dynamic properties of the system are obtained by the numerical diagonalisation of the superoperator $\mathcal{L}$. The numerical calculations are performed in a properly truncated Hilbert space, obtained by setting a maximum value $N_i^{max}$ ($i=1,2$) for the total photon occupancy per cavity. The convergence of the results versus $N_i^{max}$ is carefully checked by varying the cutoff number of photons \footnote{In this work we used $N_i^{max}=47$ for the steady-state results and $N_i^{max}=50$ for the spectral analysis of the Liouvillian.}. 

In order to study DTCs, that are collective phenomena arising in a thermodynamic limit with large photon number, it is necessary to define a proper scaling of the physical parameters, allowing to reach this limit in a controlled way. In this work we consider the thermodynamic limit obtained by letting the interaction strength $U \to 0$ and the driving amplitude $F \to \infty$ in Eq. \eqref{equa:master equation}, while keeping constant the product $UF^2$. This approach has already been used to study not only DTCs \cite{Lledo19}, but also the dissipative phase transitions in photonic system of finite size \cite{CasteelsFazio17,Casteels17,Minganti18,Hwang16,Puebla17,Hwang18}. 

In the limit of large photon occupation, the dynamics of a driven dissipative system can be generally recovered by the solution of the Gross-Pitaevskii (GP) equation \cite{CiutiRMP}, a mean field approach neglecting all fluctuations. The GP approximation is obtained from the master equation \ref{equa:master equation} assuming only coherent states for fields, $\hat{\rho} = \ket{\alpha_1,\alpha_2}\bra{\alpha_1,\alpha_2}$. The two rescaled complex fields $\alpha_1 = \sqrt{U} \langle \ao_1 \rangle$ and $\alpha_2 = \sqrt{U} \langle \ao_2 \rangle$ evolve according to the set of coupled equations:

\begin{equation}
\begin{aligned}
i\frac{\partial \alpha_1}{\partial t} & =  (-\Delta -i\kappa/2)\alpha_1 + |\alpha_1|^2\alpha_1-J\alpha_2+F\sqrt{U} \\
i\frac{\partial \alpha_2}{\partial t} & = (-\Delta -i\kappa/2)\alpha_2 + |\alpha_2|^2\alpha_2-J\alpha_1 \ .
\label{equa:GPE}
\end{aligned}
\end{equation}

The steady-state GP solutions $\alpha_{i,S}$ are obtained solving Eqs.~\eqref{equa:GPE} with the condition $i\partial_t \alpha_{i,S} = 0$. The stability of each solution can be assessed by evaluating the spectrum of linearized excitations around them. If all the frequencies of the linearized excitations have negative imaginary parts, then the corresponding solution is stable and can describe the steady state of the driven-dissipative system. Otherwise, the solution is unstable. The parametric instability happens when the frequency of the excitations presents a non-zero real part.


While the Gross-Pitaevskii formalism provides a simple approximation for the dynamics of the open quantum system, it fails in the description of the mixed character of its density matrix. To overcome this limitation, we consider another approximated scheme: the truncated Wigner approximation (TWA) method \cite{Vogel89,Opanchuk13}.  This numerical approach relies on the assumption that the equation of motion for the Wigner quasi-probability distribution function obtained from the master equation, Eq. \eqref{equa:master equation}, can be written as a Fokker-Planck equation in the limit of small non-linearities. Namely, the state of the photonic dimer can be described by two complex fields $\alpha_1(t)$ and $\alpha_2(t)$ which describe the coherence over the two modes. Their time evolution follows the stochastic differential equation

\begin{equation}
\begin{aligned}
\frac{\partial \alpha_i}{\partial t}=&-i \left[-(\Delta+i\kappa/2)+U(|\alpha_i|^2-1)\right]\alpha_i \\&-iF\delta_{i,1} + i J \alpha_{3-i} + \sqrt{\kappa/4} \chi(t) \ ,
\label{equa:TWA}
\end{aligned}
\end{equation}
where $\chi(t)$ is a normalized random complex Gaussian noise with correlators $\mean{\chi(t)\chi(t')}=0$ and $\mean{\chi(t)\chi^*(t')}=\delta(t-t')$ and describes the fluctuations arising in the quantum system because of photon losses. Each TWA trajectory corresponds to a different realization of the noise term $\chi(t)$. Therefore, the evolution of the density matrix can be recovered by averaging many trajectories obtained solving numerically the associated Langevin equation for the complex field, using stochastic Monte-Carlo techniques. In spite of its approximated nature, this approach is very useful for studying our system in regimes of high photon occupancy in the cavities, as it avoids the use of large cut-off in the number of photons per cavity.

\section{Results}
\label{sec:results}

\subsection{Mean-field analysis}

\begin{figure}
	\centering%
	\includegraphics[width=1\linewidth]{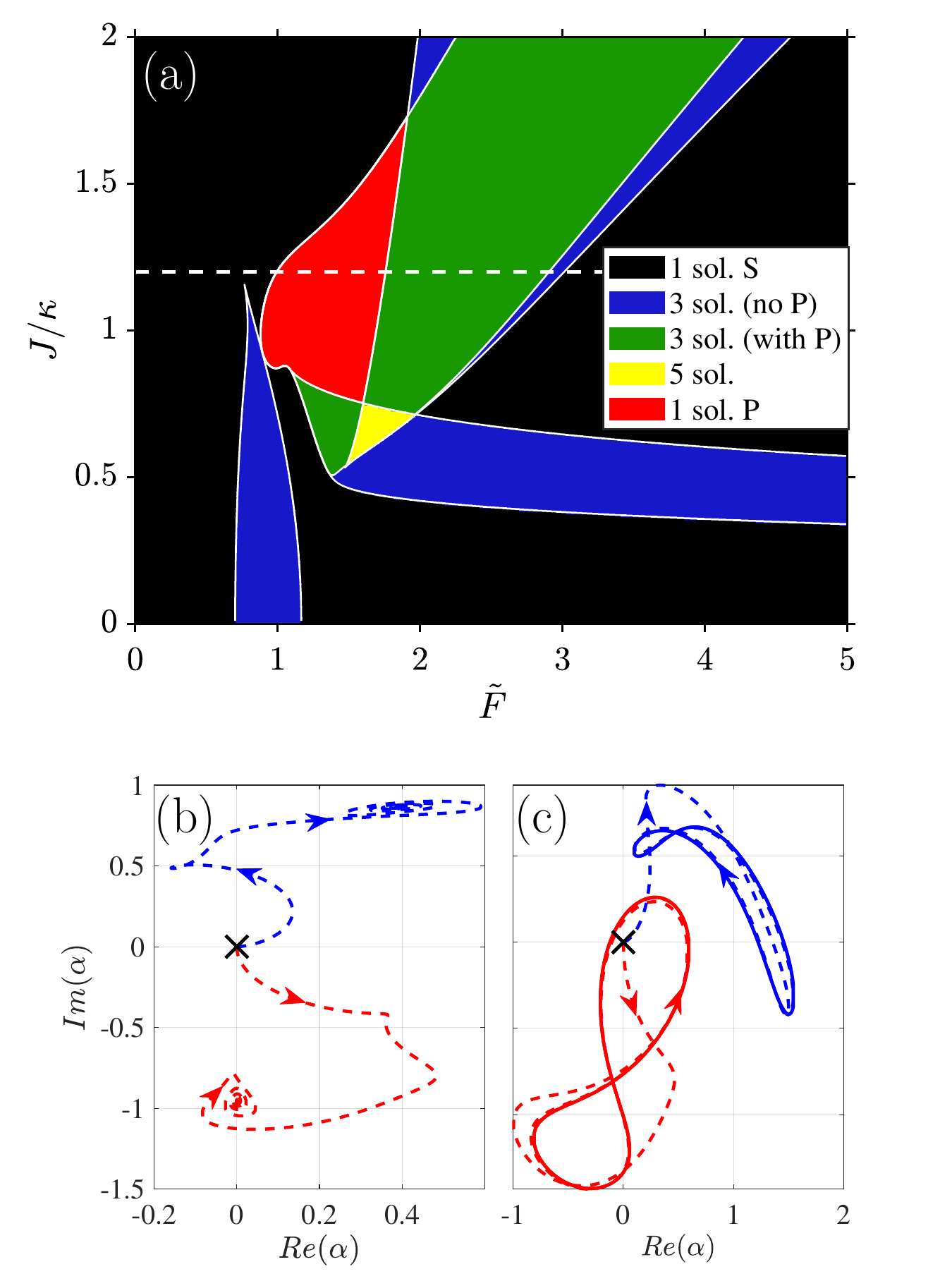}

	\caption{Panel (a): Phase diagram for the number and the nature of the GP steady state solutions as a function of $J$ and $\tilde{F}$, for the fixed value of $\Delta=2\kappa$. In the phase diagram, we distinguish the case of a single stable solution (1 sol. S), a single parametrically unstable solution (1 sol. P), three solutions (either with or without one parametrically unstable) and five solutions. The dashed line represent the value $J=1.2 \kappa$, i.e. the value of $J$ considered in the results of Sec. \ref{sec:results}.B and \ref{sec:results}.C. Panels (b) and (c): trajectories described by the mean fields $\alpha_1$ (red curve) and $\alpha_2$ (blue curve) according to the GP time evolution, for $\Delta=2\kappa$, $J/\kappa = 1.2$ and different values of $\tilde{F}$: panel (b) for $\tilde{F}=0.95$ (i.e. in the case of a single stable solution) and panel (c) for $\tilde{F}=1.5$ (i.e. in the case of a single parametrically unstable solution). The shown trajectories are obtained by numerically integrating the GP equations (Eq. \eqref{equa:GPE}) up to $t\kappa=10^3$. The arrows indicate how the fields evolve for increasing time.}
	\label{fig:phase diag}
\end{figure}

We start the discussion of our results by providing a mean-field analysis of our system, with the aim to determine the range of parameters where the DTC emerges. We can expect that the DTC phase appears whenever the GP approach predicts a unique parametrically unstable steady-state solution. For this reason, we calculate the number and the nature of the GP solutions as a function of the physical parameters $\Delta$, $J$ and $\tilde{F} = F\sqrt{U}/\kappa^{3/2}$. The results of this calculation, at the fixed value of $\Delta = 2\kappa$, are shown in the phase diagram of Fig. \ref{fig:phase diag}-(a). We clearly notice the emergence of a region where the GP approach predicts a unique parametrically unstable steady-state solution. In this regime, if we compute the time evolution of the mean fields $\alpha_i(t)$ by integrating Eqs. \eqref{equa:GPE}, choosing the vacuum as the initial condition ($\alpha_1(0) = 0$, $\alpha_2(0) = 0$), we see the emergence of limit cycles at long times, which represent the classical limit of the dissipative time crystal in the quantum system. In Fig. \ref{fig:phase diag}-(c), we plot the trajectories described by the two mean fields $\alpha_i(t)$ in the plane $\textrm{Re}(\alpha) - \textrm{Im}(\alpha)$: as the time increases, the fields do not evolve towards a steady state, but they display a periodic behavior. For comparison, in Fig. \ref{fig:phase diag}-(b), we show the time evolution of a trajectory in a regime where the GP equation predicts a single steady-state solution: in this case, each of the two mean fields $\alpha_i$ evolve towards a single point, which corresponds to the solution $\alpha_{i,S}$.

\begin{figure}
	\centering%
	\includegraphics[width=1\linewidth]{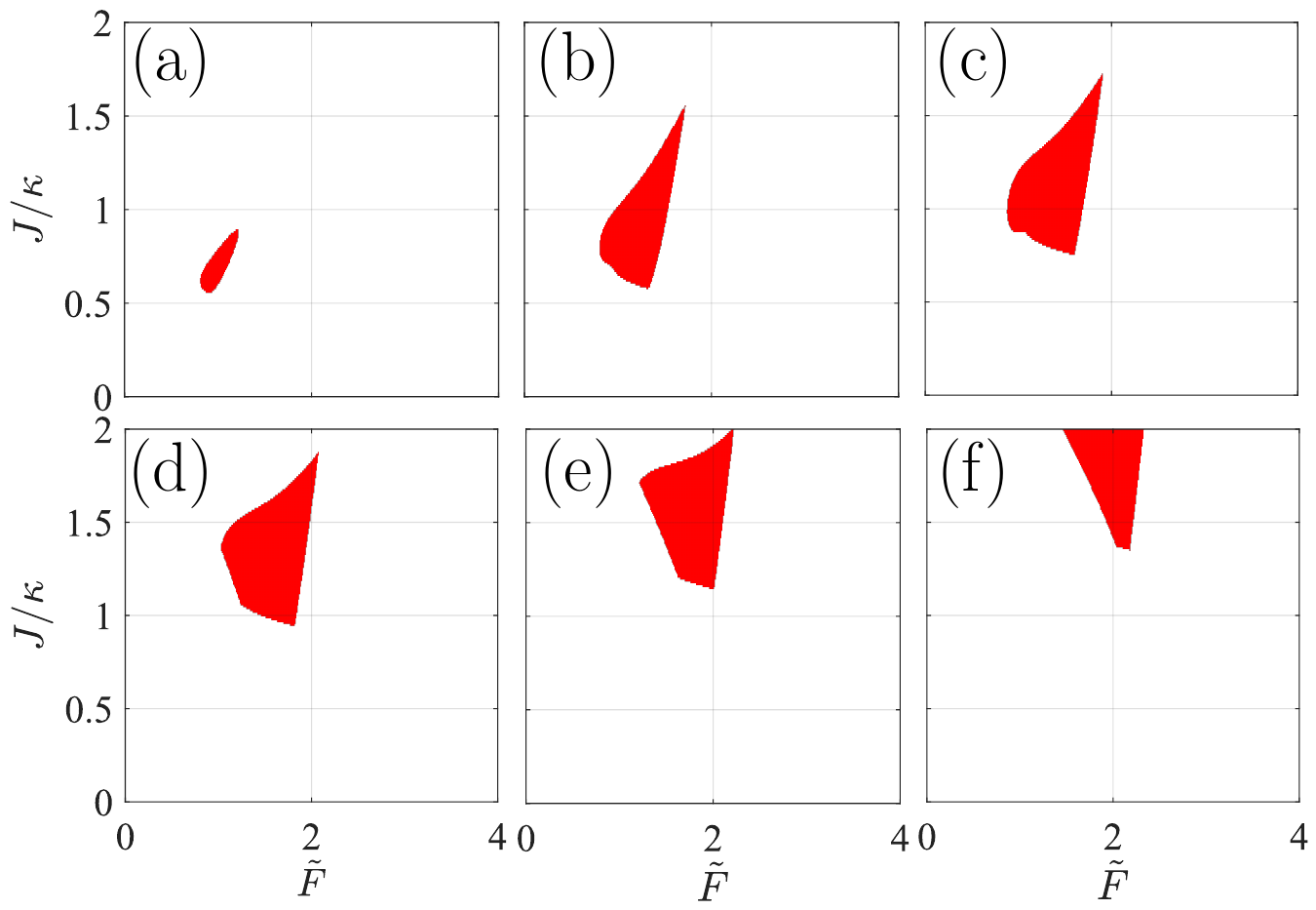}

	\caption{Region in the parameter space where GP predicts a single parametrically unstable solution: each panel show the results as a function of the parameters $J/\kappa$ and $\tilde{F}$, for different values of detuning: $\Delta/\kappa$= 1 (a), 1.5 (b), 2 (c), 2.5 (d), 3 (e), 3.5 (f).}
	\label{fig:phase diag Delta}
\end{figure}

\begin{figure}
	\centering%
	\includegraphics[width=0.8\linewidth]{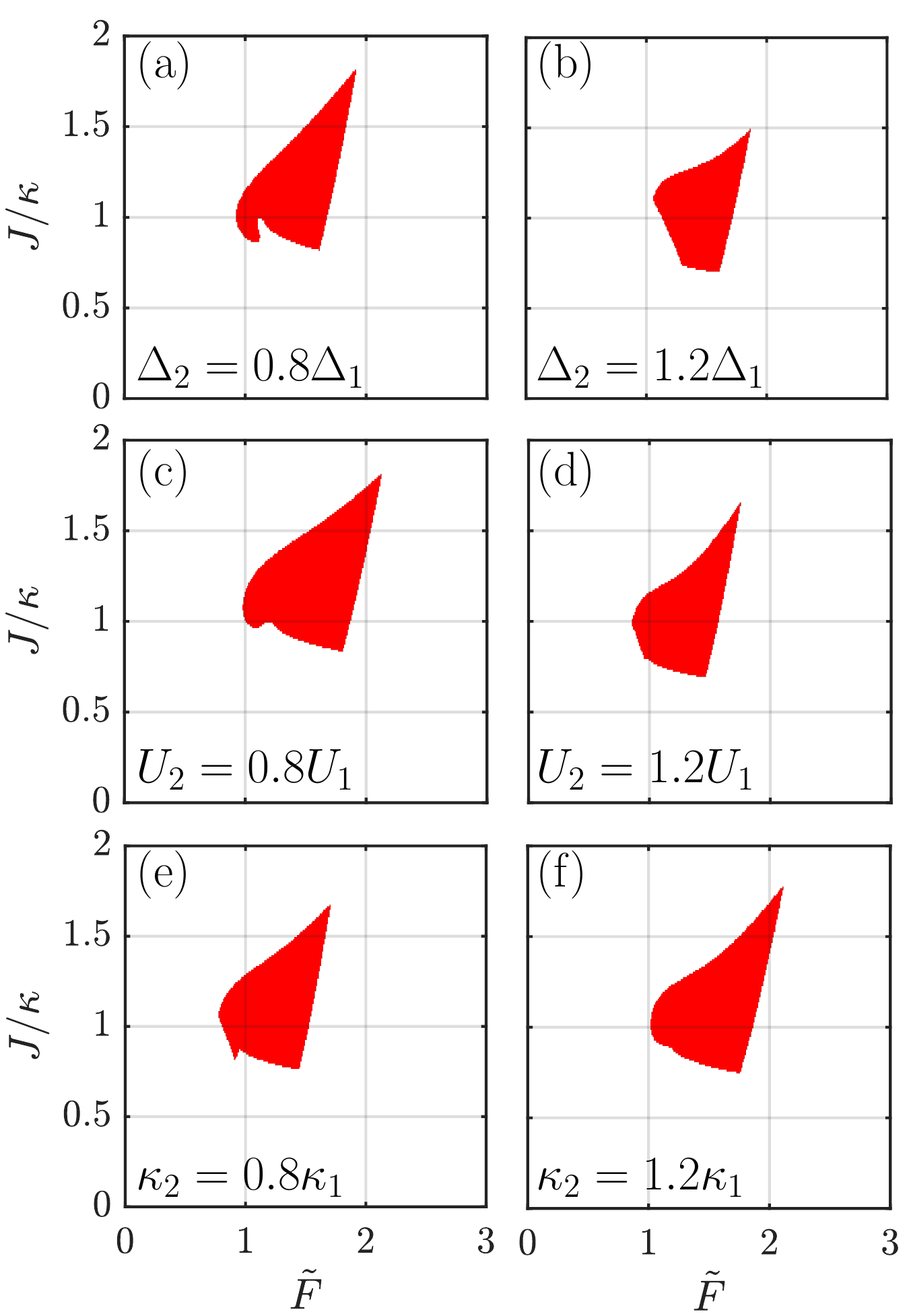}

	\caption{Region in the parameter space where GP predicts a single parametrically unstable solution, for the case of a dimer made of resonators with different detuning frequencies [panels (a-b)], nonlinearities [panels (c-d)] or loss rates [panels (e-f)]. The coefficients for the first cavity are $\Delta_1 = 2\kappa_1$ are $U_1 = \kappa_1$, $\tilde{F} = F\sqrt{U_1}/\kappa_1^{3/2}$; the coefficients for the second cavity are specified in each panel.}
	\label{fig:phase diag_differentresonators}
\end{figure}

For sake of completeness, we have derived the phase diagram for several values of the detuning $\Delta > 0$ in Fig. \ref{fig:phase diag Delta}. We notice the presence of a region with a single parametrically unstable GP solution, for all the values of $\Delta$ we have considered. Hence, it suggests that the DTC can be achieved over a finite range of values of the detuning.

In our analysis, we have considered also the case of a dimer made of two resonators with different properties. To this aim, we have solved the generalized Gross-Pitaevskii equation
\begin{equation}
\begin{aligned}
i\frac{\partial \alpha_1}{\partial t} & =  (-\Delta_1 -i\kappa_1/2)\alpha_1 + U_1 |\alpha_1|^2\alpha_1-J\alpha_2+F \\
i\frac{\partial \alpha_2}{\partial t} & = (-\Delta_2 -i\kappa_2/2)\alpha_2 + U_2|\alpha_2|^2\alpha_2-J\alpha_1 \ ,
\label{equa:GPE_DifferentResonators}
\end{aligned}
\end{equation}
which assumes different values $\Delta_1 \ne \Delta_2$ for the detuning frequencies, $U_1 \ne U_2$ for the non-linearities or $\kappa_1 \ne \kappa_2$ for the loss rates of the two modes. The results of this study are presented in fig. \ref{fig:phase diag_differentresonators}. The different panels show the region in the parameter space with a unique parametrically unstable solution for the GP steady state and indicate that the DTC phase can emerge even when the dimer is formed by two resonators with different properties, highlighting the robustness of this phase in our system.

\subsection{Steady-state properties of the quantum system}


\begin{figure}
	\centering%
	\includegraphics[width=0.9\linewidth]{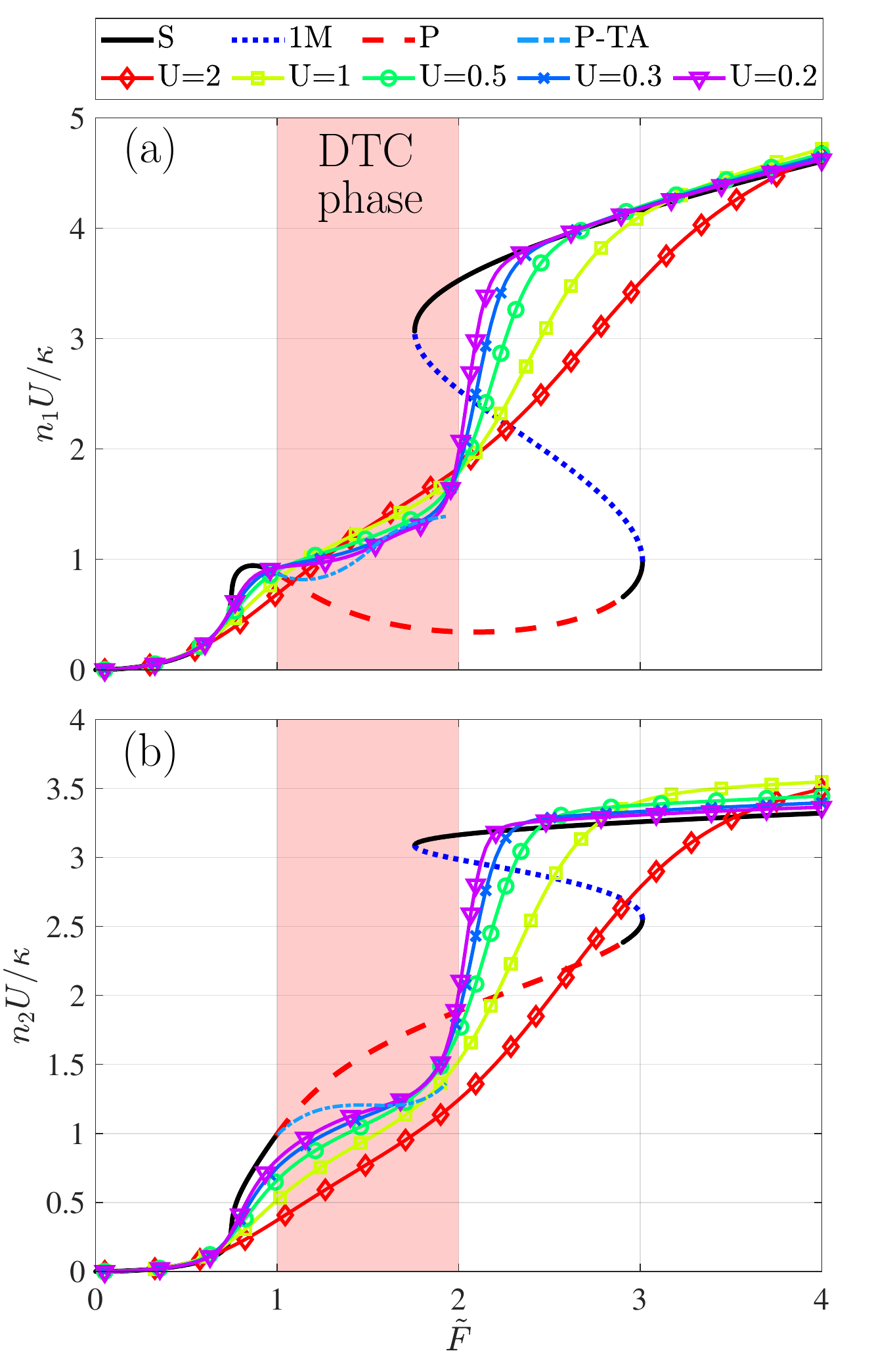}

	\caption{Rescaled steady-state expectation value of the photon occupation $n_1$ (a) and $n_2$ (b) in the two cavities, versus the rescaled pump amplitude $\tilde{F}$. The lines with markers correspond to the different values of $U$. The solid, dashed and dotted lines correspond to the predictions obtained from the steady-state solution of the GP equations, where the different line styles represent different nature of the solution: stable (solid black line), one mode unstable (blue dotted line) and parametrically unstable (red dashed line). The light blue dot-dashed line represents the prediction for the photon occupation calculated from the time-averaged density matrix $\hat{\rho}_{av}$ (Eq. \eqref{eq:TimeAverageDensityMatrix}). The shaded area indicates approximatively the DTC phase. The other Hamiltonian parameters are $\Delta/\kappa=2$, $J/\kappa=1.2$.}
	\label{fig:S shape}
\end{figure}

We now consider the properties of the system obtained within a fully quantum many-body approach.  In Fig. \ref{fig:S shape} we show the steady-state expectation values for the photon occupation $n_1 = \langle \hat{a}^\dagger_1 \hat{a}_1 \rangle$ and $n_2 = \langle \hat{a}^\dagger_2 \hat{a}_2 \rangle$ in the two cavities, as a function of the driving amplitude $F$, for different choices of the non-linearity $U$: the other Hamiltonian parameters in Eq. \eqref{equa:system Hamiltonian}, $\Delta=2\kappa$ and $J=1.2\kappa$, are chosen such that the GP equation predicts the emergence of parametric instability in the system (these parameters corresponds to the dashed line plotted in the phase diagram of Fig. \ref{fig:phase diag}). By studying the behavior of $n_1$ and $n_2$ for decreasing $U$, we can extrapolate their behavior in the thermodynamic limit and compare it with the GP prediction. We see that the mean-field approach is reliable only in the limit of small and large driving, where the GP equation predicts a unique stable steady-state solution, but it fails for intermediate values of  the rescaled driving amplitude $\tilde{F} = F\sqrt{U}/\kappa^{3/2}$. For $\tilde{F}\simeq 2$, our results show a steep increase of the photon occupancy in the two cavities as a function of $\tilde{F}$, which becomes steeper as the non-linearity $U$ decreases. This behavior suggests the emergence of a discontinuity in the thermodynamic limit, and therefore the occurrence of a first-order phase transition similar to that observed in a single cavity in regimes of optical bistability  \cite{CasteelsFazio17}. Moreover, from the results in Fig. \ref{fig:S shape} we can find a broad interval of $\tilde{F}$ values, i.e. $1 \lesssim \tilde{F} \lesssim 2$, where the expectation values computed for the quantum model do not depend strongly on $U$ (and therefore we can safely assume that the thermodynamic limit is already reached at the lowest values of $U$ achievable with our numerical approach) and are notably different from the GP predictions for the steady state. This interval of $\tilde{F}$ corresponds roughly to the range where the GP approach predicts a unique parametrically unstable steady-state solution and represents the regime where the DTC is observed (henceforth, we define this range of parameters as the DTC-phase).

Due to the emergence of limit cycles in the classical regime of the DTC-phase, it is natural to ask whether a more precise prediction for the steady state of the quantum system in the thermodynamic limit can be obtained from a time average of the dynamical solution of the GP equation. Knowing the solution for the field $\alpha_1(t)$ and $\alpha_2(t)$ over a limit cycle of period $T$, we construct the time-averaged density matrix
\begin{equation}
\hat{\rho}_{av} = \frac{1}{T} \int_{t_0}^{t_0+T} dt \ket{\alpha_1(t),\alpha_2(t)}\bra{\alpha_1(t),\alpha_2(t)} \ ,
\label{eq:TimeAverageDensityMatrix}
\end{equation}
where $\ket{\alpha_1(t),\alpha_2(t)} = \ket{\alpha_1(t)} \otimes \ket{\alpha_2(t)}$ is a coherent states on the two modes of the dimer. We compute the expectation values for the photon occupation on the two cavities $n_{i,av} = \textrm{Tr}(\hat{\rho}_{av} \hat{a}_i^\dagger \hat{a}_i)$ over this density matrix. The results for  $n_{i,av}$ as a function of $\tilde{F}$ are showed in Fig. \ref{fig:S shape} and compared with the results obtained from the steady-state solution of the master equation, Eq. \eqref{equa:master equation}. We notice that $n_{i,av}$ is in agreement with the quantum results in the thermodynamic limit.



%

\begin{figure}
	\centering%
	\includegraphics[width=1\linewidth]{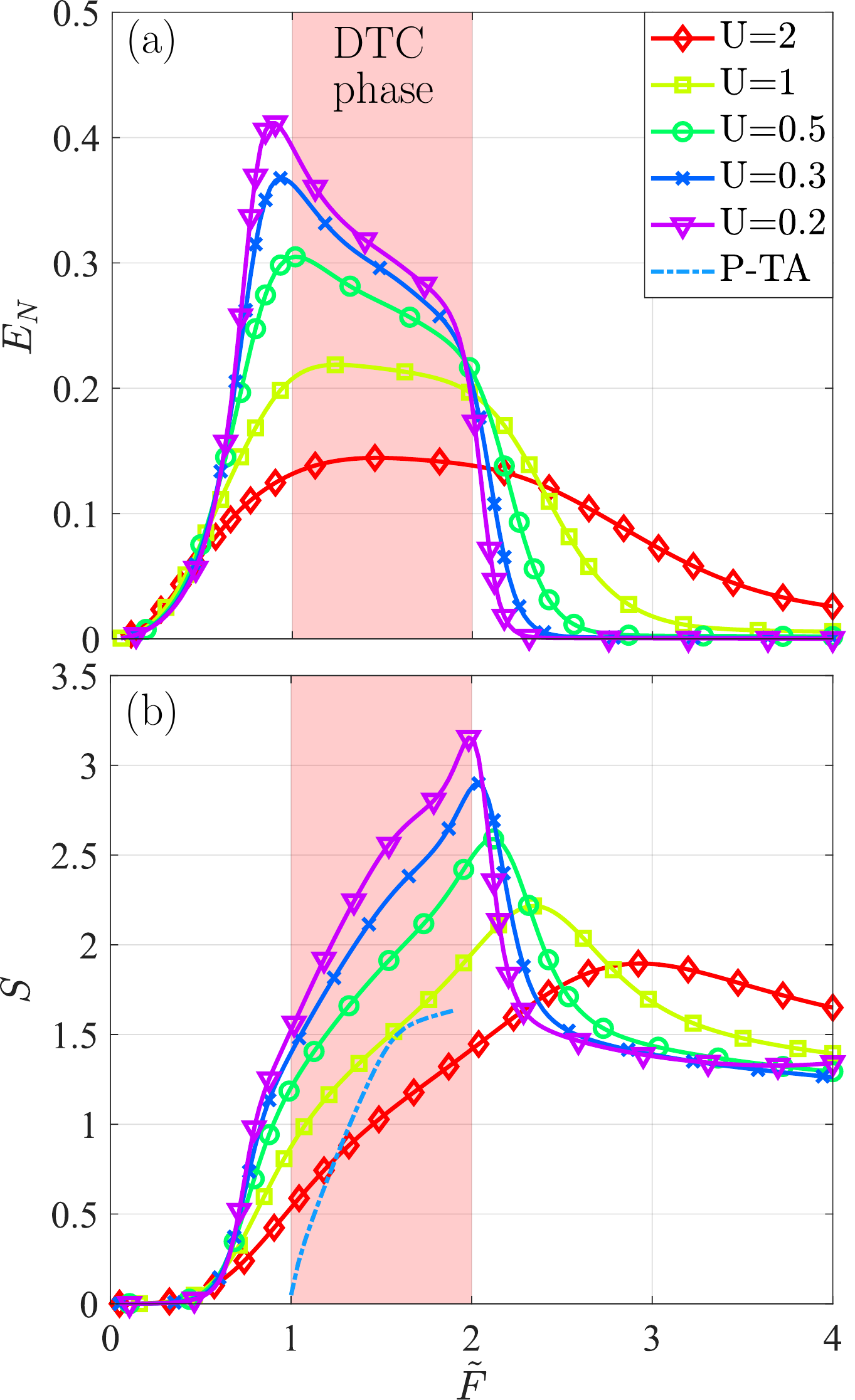}

	\caption{Logarithmic negativity $E_N$ (a) and Von Neumann Entropy $S$ (b) versus the rescaled pump amplitude and for different values of the non-linearity $U$.  The shaded area indicates approximatively the DTC phase. In panel (b), the light blue line represents the prediction for the von-Neumann entropy calculated from the time-averaged density matrix $\hat{\rho}_{av}$ (Eq. \eqref{eq:TimeAverageDensityMatrix}). The other Hamiltonian parameters are $\Delta/\kappa=2$, $J/\kappa=1.2$.}
	\label{fig:NegEntro}
\end{figure}

However, the density matrix $\hat{\rho}_{av}$ is not able to give a complete description of the steady state $\hat{\rho}_{ss}$ of the quantum system, as evidenced by the results for the logarithmic negativity $E_N$ and for the von-Neumann entropy, which are shown in Fig. \ref{fig:NegEntro}. In Fig. \ref{fig:NegEntro}-(a), we show $E_N = \log_2(||\op{\rho}_{ss}^{\Gamma_1}||_1)$, where $\op{\rho}_{ss}^{\Gamma_1}$ indicates the partial transpose with respect to the degrees of freedom of the second cavity and $||.||_1$ the trace norm, as a function of $\tilde{F}$. We can see that, in the DTC-phase, $E_N > 0$ and increases for decreasing $U$, showing the presence of entanglement, which is absent per definition in $\hat{\rho}_{av}$ (Eq. \eqref{eq:TimeAverageDensityMatrix}). The von-Neumann entropy $S=-\textrm{Tr}(\op{\rho}_{ss}\ln(\op{\rho}_{ss}))$, displayed in Fig. \ref{fig:NegEntro}-(b), shows instead the mixed character of the steady state, arising because of the classical fluctuations due to the photon losses from the cavities. We see that $S$ assumes large values for $1 \lesssim \tilde{F} \lesssim 2$ and increases for decreasing $U$. We notice that the classical prediction $S_{av}=-\textrm{Tr}(\op{\rho}_{av}\ln(\op{\rho}_{av}))$, also shown in Fig. \ref{fig:NegEntro}-(b), does not agree with the quantum results in the thermodynamic limit. The analysis of $E_N$ and $S$ confirms the important role played by fluctuations (both quantum and classical) in the steady state of our system and therefore the inaccuracy of the GP approach in the description of the DTC-phase.

\subsection{Quantum dynamics}

\begin{figure}
	\centering%
	\includegraphics[width=1\linewidth]{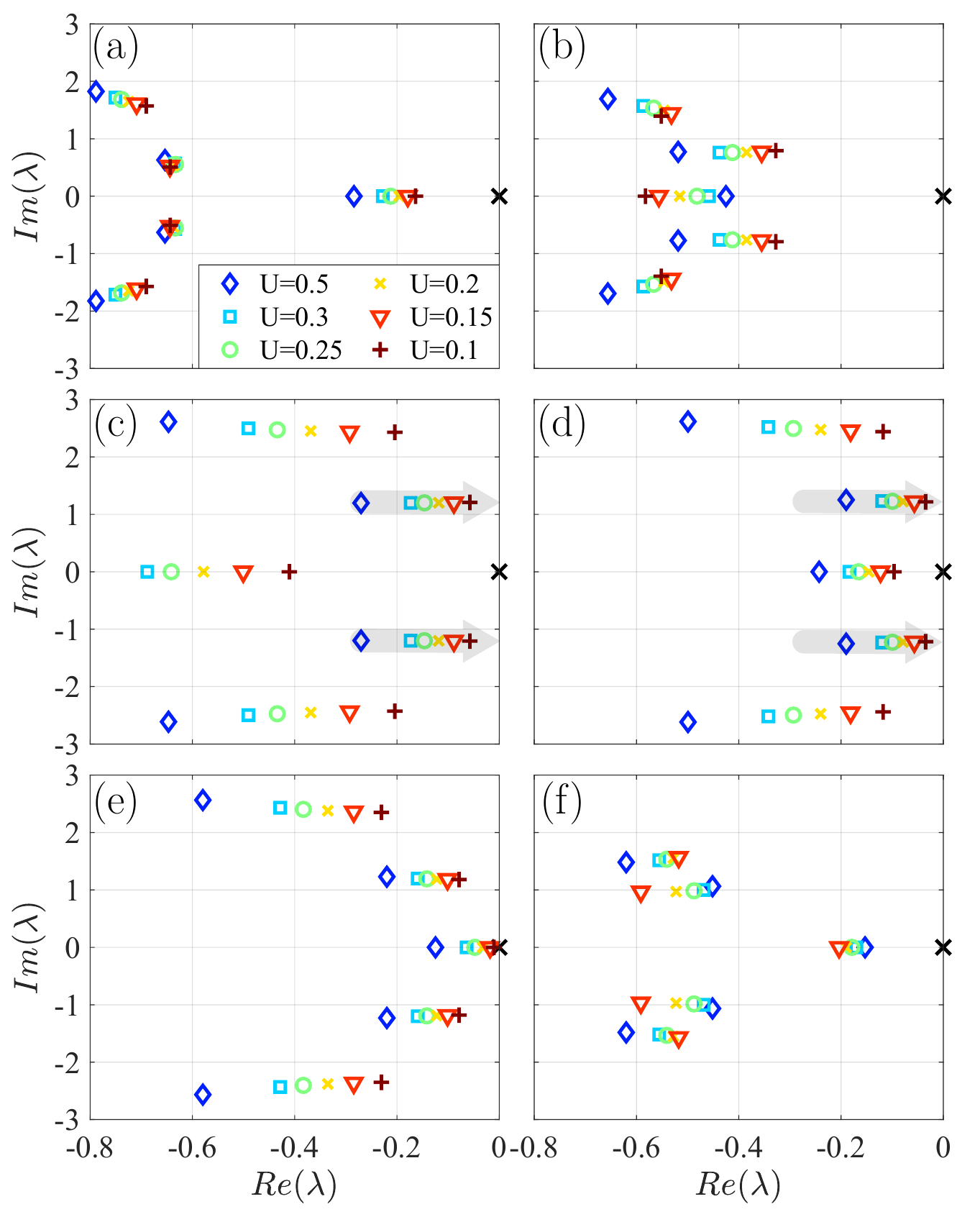}

	\caption{Eigenvalues $\lambda$ of the Liouvillian superoperator, plotted in units of $\kappa$. Each panel shows the eigenvalues with largest real part for a given pump amplitude $\tilde{F}$ and different values of non-linearity $U$, corresponding to different colors and marker types. The black cross indicates the unique steady state. The different pump amplitudes corresponding to each panels are: $\tilde{F}=0.8$ (a), $\tilde{F}=1.0$ (b), $\tilde{F}=1.5$ (c), $\tilde{F}=1.8$ (d), $\tilde{F}=2.0$ (e) and $\tilde{F}=2.5$ (f). The other parameters are $\Delta/\kappa=2,\; J/\kappa=1.2$.}
	\label{fig:Liouvillian spectrum Re/Im}
\end{figure}
In order to reveal the emergence of a DTC in the considered system, we study the dynamical properties by computing the spectrum of the eigenvalues $\lambda_j$ of the Liouvillian. This is performed by numerically diagonalizing the superoperator $\mathcal{L}$ defined in Eq. \eqref{equa:master equation}. In Fig.  \ref{fig:Liouvillian spectrum Re/Im}, we show the spectrum of the eigenvalues of $\mathcal{L}$ with largest real part, for different values of $\tilde{F}$ and $U$. Outside of the DTC-phase [see Fig. \ref{fig:Liouvillian spectrum Re/Im}-(a) for $\tilde{F} = 0.8$ and Fig. \ref{fig:Liouvillian spectrum Re/Im}-(f) for $\tilde{F} = 2.5$], the eigenvalue with smallest absolute value is purely real, independently of the value of $U$. In this regime, the dynamics of the dissipative system at long time is characterized by an exponential decaying towards the steady state. At $\tilde{F}=1$ [Fig. \ref{fig:Liouvillian spectrum Re/Im}-(b)], the onset of long-lived oscillation at small $U$ is revealed by the fact that the eigenvalue $\lambda_1$ with largest real part has a finite imaginary part. We also see, in this case, that the Liouvillian gap $\Lambda = |\textrm{Re}(\lambda_1)|$ decreases for decreasing $U$. 

The typical Liouvillian spectrum in the DTC-phase is shown in Fig. \ref{fig:Liouvillian spectrum Re/Im}-(c) ($\tilde{F} = 1.5$) and Fig. \ref{fig:Liouvillian spectrum Re/Im}-(d) ($\tilde{F} = 1.8$). From these plots, we clearly notice the presence of eigenvalues which, when $U \to 0$, have a vanishing real part and finite imaginary part. This means that the time scale of the relaxation dynamics (which is determined by the inverse of the Liouvillian gap $1/\Lambda$) becomes increasingly long when approaching the thermodynamic limit. Even though the Lindblad master equation Eq. \eqref{equa:master equation} predicts the existence of a time-independent steady state, the evolution of the density matrix is characterized by long lived oscillations: indeed, according to the spectral decomposition of the density matrix \cite{Minganti18}, \begin{equation}
\hat{\rho}(t) = \hat{\rho}_{ss} + \sum_j c_j(0) e^{\lambda_j t} \hat{\rho}_j \ ,
\end{equation}
where $\hat{\rho}_j$ are the eigenmatrices of the Liouvillian superoperators associated to the eigenvalues $\lambda_j\neq 0$, and $c_j(0)$ are the components of the initial density matrix $\hat{\rho}(0)$ over the different $\hat{\rho}_j$. While all the components having $\lambda_j$ with sizeable real part decay rapidly, those with $|\textrm{Re}(\lambda_j)| \ll |\textrm{Im}(\lambda_j)|$ will give rise to long lived oscillations in $\hat{\rho}(t)$.

The results in Fig. \ref{fig:Liouvillian spectrum Re/Im}-(c,d) suggest also that the imaginary part of the eigenvalues with vanishing real part are integer multiples of a fundamental frequency. The two features, i.e. the gapless Liouvillian spectrum and the imaginary eigenvalues of the low excitations described by bands separated by the same frequency, are the key elements of a DTC, as also pointed out in Ref. \cite{Iemini18}.

Finally, for $\tilde{F} = 2.0$ [Fig. \ref{fig:Liouvillian spectrum Re/Im}-(e)], we notice that the eigenvalue with largest real part is purely real, signaling the disappearance of the long-lived oscillation of the DTC-phase. Moreover, we can notice also that this eigenvalue goes to zero in the thermodynamic limit: this behavior can be associated to the closing of the Liouvillian gap in the vicinity of a critical point, and hence supports the evidence for a first-order dissipative phase transition \cite{Minganti18}, as already indicated by the results in Fig. \ref{fig:S shape}.

\begin{figure}

\centering%
\includegraphics[width=1\linewidth]{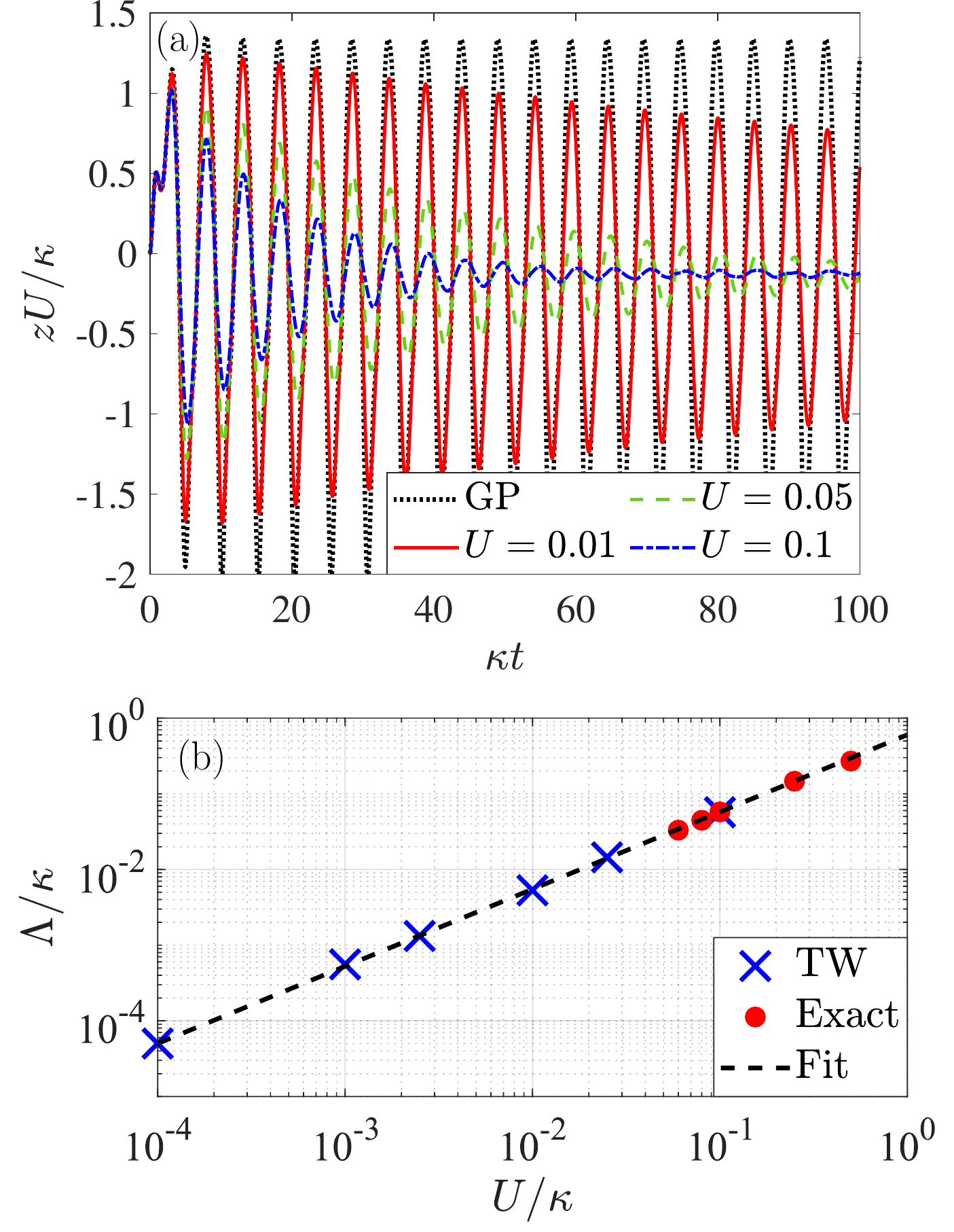}
	\caption{(a): Time evolution of the rescaled population difference $z$ between the two cavities for $\tilde{F}=1.5$, $\Delta/\kappa=2,\; J/\kappa=1.2$ and different values of $U$. The results are obtained by averaging over $10^5$ stochastic trajectories obtained with TWA. (b): Liouvillian gap $\Lambda$ as a function of the non-linearity $U$ for $\tilde{F}=1.5$, $\Delta/\kappa=2,\; J/\kappa=1.2$. The different symbols refer to different methods to extract the value of $\Lambda$ (exact diagonalization or fit of TWA results for $z(t)$). The dashed line represents a power-law fit of the data.}
	\label{fig: t evol TW}
\end{figure}

The occurrence of a DTC in our system is further supported by a study of the dynamics with the Truncated Wigner approximation (TWA) \cite{Vogel89}. In Fig. \ref{fig: t evol TW}-(a), we show the TWA results for the time evolution of the population difference between the two cavities, $z = \langle \hat{a}_1^\dagger \hat{a}_1 - \hat{a}_2^\dagger\hat{a}_2 \rangle$, for the value of $\tilde{F} = 1.5$ inside the DTC-phase, having chosen the vacuum as the initial condition. The oscillating character of the dynamics is evident for all the values of the non-linearities considered and persists on a time scale which is large with respect to the inverse loss rate $1/\kappa$. By comparing the curves obtained for different values of $U$, we can see that the damping of the oscillation becomes smaller for decreasing $U$, but their period is almost independent. These results confirm what already observed in the analysis of the Liouvillian spectrum: when approaching the thermodynamic limit, the Liouvillian gap goes to zero, as indicated by the slowing down of the exponential decay of the oscillation; instead its imaginary part, which is related to the period of the oscillations, remains finite. The numerical estimates for the Liouvillian gap $\Lambda$ and for the relevant frequencies of the oscillation can be extracted by fitting the curves $z(t)$ at long times with a sum of exponentially damped sine functions. The behavior of $\Lambda$ as a function of $U$ for $\tilde{F} = 1.5$ is shown in Fig. \ref{fig: t evol TW}-(b). First of all, we notice that the values extracted from the fit of $z(t)$ are in good agreement with the results obtained by the exact diagonalization of the Liouvillian, confirming the validity of the TWA in this regime of small non-linearities. Furthermore, the results show a power law behavior $\Lambda \sim U^\eta$, with $\eta = 1.02 \pm 0.03$, indicating that the Liouvillian gap closes in the thermodynamical limit. Concerning the oscillatory dynamics of the system in the DTC-phase, the frequencies extracted from the fit of $z(t)$ at the largest $U$ correspond exactly to the imaginary part of the Liouvillian eigenvalues shown in Fig. \ref{fig:Liouvillian spectrum Re/Im}-(c). When $U$ decreases, it becomes apparent that more frequency components contribute to the oscillation of $z(t)$. All the frequencies extracted from the fit are integer multiples of the same fundamental frequency: this picture strongly supports the presence of a discrete set of equally spaced Liouvillian eigenvalues, that is a sufficient condition to have a persistent non-stationarity in the dynamics of the open system \cite{Buca2019Arxiv}. A rigorous proof of this spectral structure in the thermodynamic limit is however beyond the scope of this work.

\begin{figure}
	\centering%
	\includegraphics[width=0.85\linewidth]{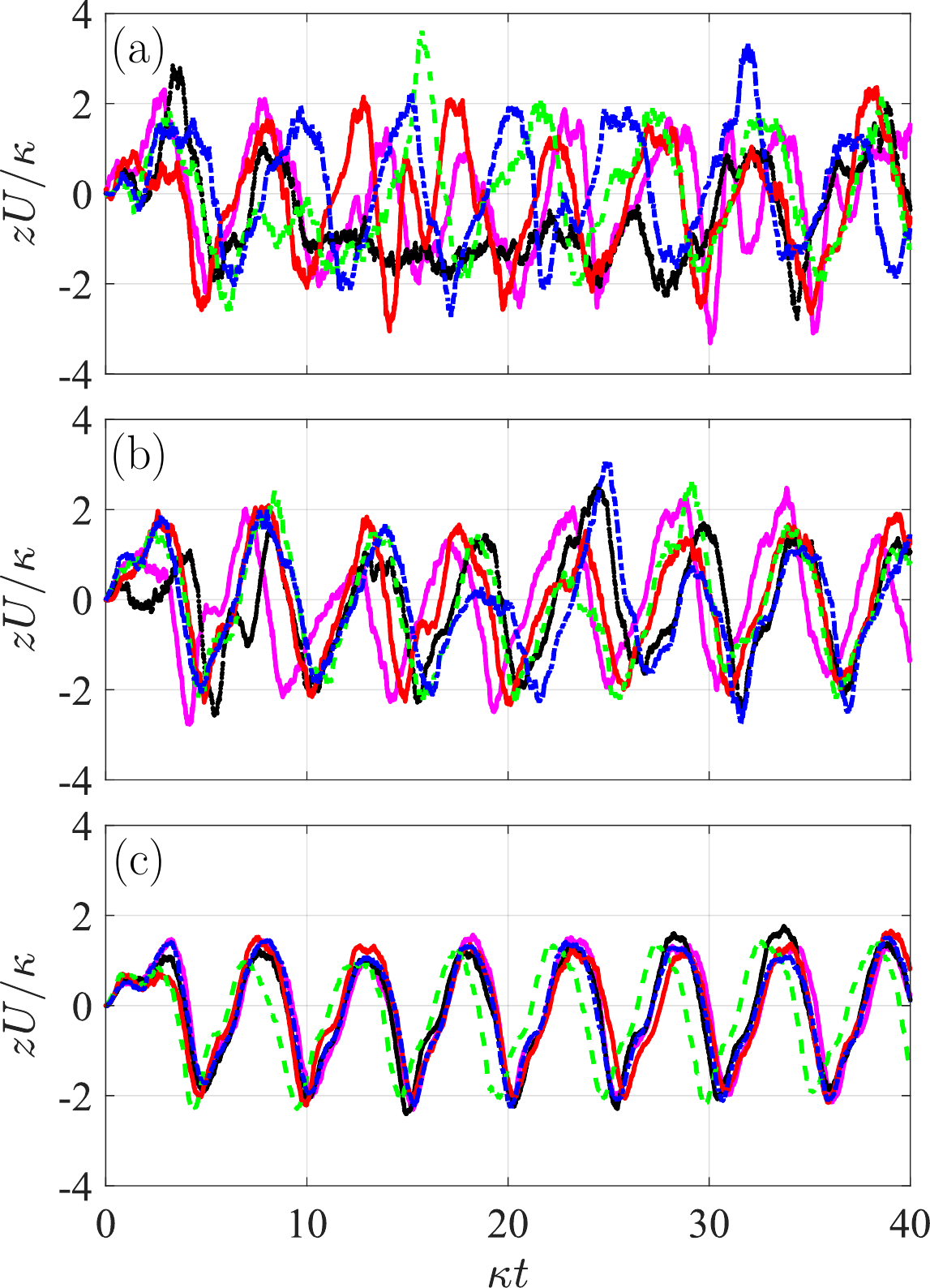}

	\caption{Time evolution of the population difference between the two cavities obtained for 5 different TWA trajectories, starting from the vacuum at $t = 0$. The different panels show the results for different non-linearities: $U=0.1$ (a), $U=0.05$ (b) and $U=0.01$ (c). The other parameters are $\tilde{F}=1.5$, $\Delta/\kappa = 2$ and $J/\kappa = 1.2$ }
	\label{TW single traj U}
\end{figure}

In Fig. \ref{fig: t evol TW}, we show the quantity $z(t)$ obtained from the GP equation, when taking the vacuum as initial condition at $t=0$. The comparison with the TWA results shows that the fluctuations do not affect the frequency of the oscillations. Fluctuations are instead only responsible for random relative phase shifts among single TWA trajectories, resulting in the damping of oscillations at long times. This last observation is verified by comparing the behavior of individual TWA trajectories. In Fig. \ref{TW single traj U}, we show the evolution of the population difference between the two cavities, over five TWA trajectories, for different values of the non-linearity $U$ and for the value of $\tilde{F} = 1.5$, inside the DTC-phase. For all the values of $U$, we notice that most of the single trajectories present an oscillating behavior, which persists over a longer time interval as the non-linearity decreases. 
From this analysis, we can deduce that the fluctuations induced by the noise term $\chi$ in Eq. \eqref{equa:TWA} do not suppress the oscillating character of the trajectories, but induce a certain dephasing among them, which results in the damping towards the steady-state expectation value when the results of the single trajectories are averaged (See Fig. \ref{fig: t evol TW}).

To have a better understanding of how the fluctuations influence the dynamics of the system in the DTC-phase, we show in Fig. \ref{fig:TW U t} the distribution of the fields $\alpha_i(t)$ over a set of 12000 TWA trajectories for different times $t$ and non-linearities $U$. At short times, the distribution is a Gaussian centered around the GP solution. At longer times, the effect of the noise is to spread the distribution of $\alpha_i$ along the limit cycles defined from the GP equation (showed in Fig.~\ref{fig:phase diag}-(c)). Thus, even though a single TWA trajectory does not reach a steady state but presents an oscillating character similar to that of the GP parametrically unstable solution, the full distribution becomes stationary for long time, showing the emergence of a steady state. The time interval needed to reach the steady state becomes larger when $U$ decreases.
\newpage
\begin{widetext}
	\onecolumngrid
	\begin{figure}
		\centering%
		\includegraphics[width=1\textwidth]{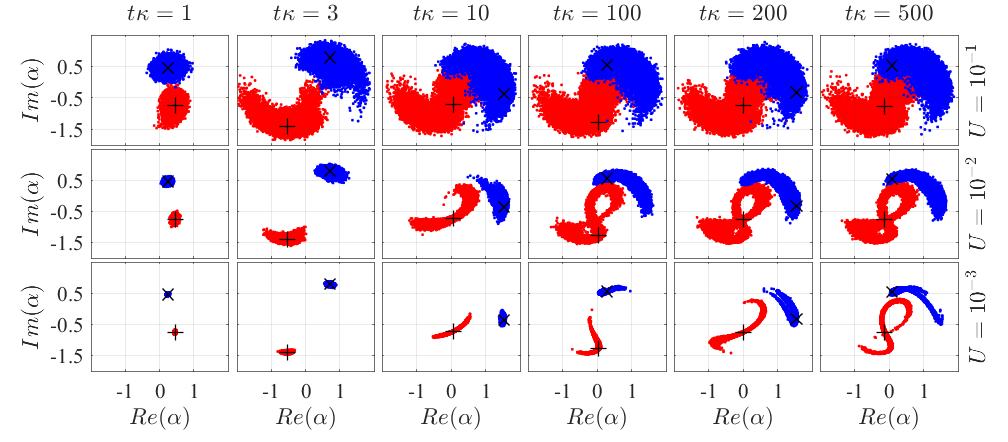}

		\caption{Distribution of the TWA fields $\alpha_1$ (in red) and $\alpha_2$ (in blue) in phase space at different times $t$ and non-linearities $U$. The black markers give the GP solution for the first (+) and second ($\times$) cavity for the given time. The distributions are obtained from the realization of $1.2 \times 10^4$ TWA trajectories.}
		\label{fig:TW U t}
	\end{figure}
	\twocolumngrid
\end{widetext}

\section{Conclusions}
\label{sec:conclusions}

In conclusion, we have provided strong evidences of the occurrence of a dissipative time crystal in a simple driven-dissipative system of two coupled non-linear optical resonators, under general conditions which do not rely on the presence of symmetries. The DTC phase arising over a wide range of parameters is characterized by spontaneous long lived oscillations of the system observables under continuous-wave driving, large fluctuations and non-classical correlations. The scheme we propose can be easily realized with current experimental technologies, such as superconducting circuits \cite{Eichler14} or semiconductor micropillars \cite{Abbarchi2013,Galbiati12,Lagoudakis2010,Rodriguez2016}, which have already been used for the investigation of other collective phenomena in open quantum system. The emergence of a DTC in an optical dimer is directly related to the physics of Kerr solitons, for which the quantum properties of the radiation field are yet to be fully understood. The present study is an important step toward the characterization of quantum correlations and entanglement in Kerr-soliton systems, opening the way to the design of optical devices for the generation of non-classical light.


\section{Acknowledgments}
We would like to thank Fabrizio Minganti and Wouter Verstraelen for useful discussions. We acknowledge support from the Swiss National Science Foundation through Project No. 200021\_162357 and 200020\_185015.

\bibliographystyle{apsrev4-1}
\bibliography{ParamInstabilitiyDimer_final}

\end{document}